\documentstyle[prb,preprint,aps]{revtex}
\begin{document}
\draft
\title{Transport in  $\alpha$-Sexithiophene  Films}
\author{M. W. Wu and E. M. Conwell}
\address{Center for Photoinduced Charge Transfer, Chemistry Department,
University of Rochester, Rochester, New York 14627, USA\\
and Xerox Corporation, Wilson Center for Technology, 114-22D, Webster,
New York 14580, USA$^*$}
\maketitle

\begin{abstract}
The field-effect mobility of hole polarons in $\alpha$-sexithiophene,
measured in thin film transistors, was shown to be well fitted
by Holstein's small polaron theory.  Unfortunately, Holstein's
formulation is based on an integral that does not converge.
We show  that the data are well fitted by a theory of polaron
transport that was successful in accounting for mobility in
molecular crystals of naphthalene.
\end{abstract}
\newpage

The prospect of field-effect transistors, FETs, made of easily
processible thin organic films has stimulated  considerable research
effort.  The most successful devices to date have been made of
oligomeric rather than polymeric material, an outstanding one
being $\alpha$-sexithiophene, $\alpha$-6T.\cite{hor1,do}
From the device theory and  operation it is possible to deduce a
field-effect mobility, $\mu_{FET}$, for holes moving in the
active part of the device, the $\alpha$-6T layer just above
the gate dielectric.  Recent measurements of mobility made in
the space-charge limited current regime have been in good agreement
with $\mu_{FET}$ obtained in a long channel device,\cite{hor1}
indicating that the device measurements can be considered to represent
the mobility reasonably well.  For convenience of notation
we will drop the subscript in referring to the device measurements.

The $\alpha$-6T films for the FETs are usually made by vacuum
evaporation onto the gate dielectric, the latter frequently
being SiO$_2$ or MgF$_2$. This results in a polycrystalline film which
is, however, well oriented in that the long axis of the $\alpha$-6T
molecules stands almost perpendicular to the subtract.\cite{ga,lo}
The transport direction in the FETs, parallel to the substrate,
is then essentially perpendicular to the molecular chains.  Mobility
as a function of temperature has been measured in $\alpha$-6T FETs by
Horowitz {\em et al}.,\cite{hor2} Waragai {\em et al}.\cite{wa}
and Torsi {\em et al}.\cite{to}  In the two former cases the measurements
were taken down to 100 and 77\ K, respectively.  Over this range mobility
was seen to increase with increasing temperature.  Waragai et al.
attributed the transport to thermally activated hopping of polarons
among the thiophene molecules. (Actually they used a derivative of
$\alpha$-6T, dimethyl sexithiophene.)  Horowitz et al. analyzed the data
down to 150 K in terms of free carriers, specifically holes in the
valence band, undergoing multiple thermal trapping and release
involving shallow traps.  Below 150\ K they attributed the mobility to
the free carriers hopping among deep traps.  Such a model could
never account  for the increase in mobility with decreasing temperature
found by Torsi et al. below $\sim$ 50\ K.\cite{to} (See Fig.\ 1.)
The model would be appropriate for carriers moving in wide bands of
the kind found in three-dimensional semiconductors such as silicon, but
not for carriers moving in the narrow bands characteristic of a
molecular crystal, or the polycrystalline array of molecular crystals
that $\alpha$-6T samples generally consist of.  For these cases it
is essential to include polaronic effects.

Torsi {\em et al}. showed that their data are well fitted by Holstein's
small polaron theory based on his Molecular Crystal Model.\cite{hol} In
this model the electron (or hole) wavefunction is limited to a
single site.  The electron or hole becomes a small polaron, still
limited to a single site, due to its coupling to the vibrations
of a diatomic molecule at the site.  It is well known that an
excess electron or hole on a conducting polymer or on a conjugated
molecule such as naphthalene or $\alpha$-6T forms a polaron that is
spread over a considerable number of sites, $\sim$ 20 if the conjugation
length allows.  Although Holstein's calculations are readily generalized
to such a polaron, and to the vibrations being the usual acoustic
and optical modes of a lattice,\cite{con1} they are fatally flawed by the
fact that their essential result, the integral obtained for
the probability of a polaron moving or hopping from site to site,
does not converge.\cite{de,emin}
Holstein obtains finite results for the high
temperature (hopping) region and the low temperature (coherent
motion) region by approximating the integral for each case with a
leading term.  Nevertheless, many of the ideas developed by Holstein
and some predecessors form the basis for later theories of polaron
transport.

A form of the theory that was successful in describing mobility
over a wide range of temperatures for naphthalene crystals
starts from a simple Hamiltonian used by many workers in this field:
\begin{eqnarray}
H&=&\sum_m\epsilon_ma^\dagger_ma_m+\sum_{m,n}V_{m-n}a^\dagger_ma_n+
\sum_{\bf q}\hbar\omega_{\bf q}(b_{\bf q}^\dagger b_{\bf q}+
\frac{1}{2})\nonumber\\
&&+N^{-1/2}\sum_{m,{\bf q}}\hbar\omega_{\bf q}g_{\bf q}\exp(i{\bf q}
\cdot{\bf R}_m)(b_{\bf q}+b_{-{\bf q}}^\dagger)a_m^\dagger a_m\;,
\end{eqnarray}
where $a_m^\dagger$ creates an electron with energy $\epsilon_m$
at a site $m$ in the crystal of $N$ sites, $b_{\bf q}$ creates a
phonon with wave vector ${\bf q}$ and frequency $\omega_{\bf q}$ ,
$V_{m-n}$ is the intersite matrix element between sites $m$ and $n$,
$g_{\bf q}$ is the dimensionless electron-phonon coupling constant
and ${\bf R}_m$ is the lattice vector that locates site $m$.
The Hamiltonian is transformed to the polaron picture by a unitary
transformation,\cite{comm} resulting in
\begin{eqnarray}
H&=&\sum_m\left\{\epsilon_m-N^{-1}\sum_{\bf q}|g_{\bf q}|^2\hbar
\omega_{\bf q}\right\}A_m^\dagger A_m+\sum_{\bf q}\hbar\omega_{\bf q}
(B_{\bf q}^\dagger B_{\bf q}+\frac{1}{2})\nonumber\\
&&\mbox{}+\sum_{n,m}V_{n-m}\theta_n^\dagger\theta_m A_n^\dagger A_m
+N^{-1}\sum_{n,m,{\bf q}}|g_{\bf q}|^2\hbar\omega_{\bf q}
\exp[i{\bf q}\cdot({\bf R}_n-{\bf R}_m)]A_n^\dagger A^\dagger_m
A_nA_m\;,\\
\theta_{m}&=&\exp\left[N^{-1/2}\sum_{\bf q}g_{\bf q}
\exp(i{\bf q}\cdot{\bf R}_m)(B_{\bf q}-B_{-{\bf q}}^\dagger)\right]\;,
\end{eqnarray}
where $A_m$ creates a polaron at site $m$ and $B_{\bf q}^\dagger$
creates a phonon of wave vector ${\bf q}$ that is associated with the
motion of the ions about their displaced equilibrium positions.

The Hamiltonian (2), (3) was used in the calculation of polaron
mobility in naphthalene by Kenkre {\em at al.}\cite{ken}
The formulation of KADD includes disorder, the site energy
$\epsilon_m$ being taken as a randomly fluctuating quantity.  The
fluctuations can arise from crystal defects, which are surely
present in these materials.  It is usual to represent $\epsilon_m$ by
a Gaussian or Lorentzian distribution of width $\hbar\alpha$.  The
latter was chosen by KADD.  It was specified that $\alpha$ must be small
compared to $\omega_{\bf q}$; if $\alpha$ were not small a different
approach to the problem would be necessary.

As was done by Holstein and many others, KADD assumed that the hole
interacts with only one phonon branch of mean energy $\hbar\omega_0$
and width $\hbar\Delta_i$ in the $i$th direction. The significance
of this assumption will be discussed below.  With these
simplifications KADD obtained the mobility $\mu_{ii}$ for field and
current along the $i$th principal axis direction:
\begin{eqnarray}
\label{mu}
\mu_{ii}&=&2(ea_i^2)(k_BT)^{-1}|V_i/\hbar|^2\exp[-2g_i^2\coth(\frac{1}{2}
\beta\hbar\omega_0)]\nonumber\\
&&\times\int_0^\infty dt e^{-\alpha t}I_0\{2g_i^2\mbox{csch}(\frac{1}{2}
\beta\hbar\omega_0)|J_0(\Delta_1t)J_0(\Delta_2t)J_0(\Delta_3t)|
[1+J_1^2(\Delta_it)/J_0^2(\Delta_it)]^{1/2}\}\;.
\end{eqnarray}
where $\beta=1/k_BT$, the $g_i$ are dimensionless electron-phonon
coupling constants as defined in Eq.\ (1), the $a_i$ are lattice
constants and the $V_i$ the nearest-neighbor transfer integrals
along the $i$th direction.  The bandwidth $W$ is taken as usual as $4V_i$.
It is seen that $\mu_{ii}$ is similar to the result of Holstein's
theory, but it converges due to the factor $e^{-\alpha t}$, representing
disorder, in the integral.

It is clear that many phonon branches, not only one, have
significant interaction with the current carrier, whether
electron or hole.  For example, when a carrier jumps onto or off a
chain, the well known distortion that occurs involves a number of
phonon frequencies and more than one phonon branch.  This was
calculated explicitly for the case of polyacetylene.\cite{con2} A similar
polaronic effect must  occur when a hole  jumps onto or off
an $\alpha$-6T molecule.  Of course the distortions are not large, but
this represents a necessary energy expenditure that affects the
hopping rate.  A much bigger effect, however, can be caused by phonons
that change the overlap of the molecules between which the polarons jump.
For example, consider nearest neighbor hopping along that direction
in an $\alpha$-6T crystallite perpendicular to the long axis of
the molecules where nearest neighbors are tilted at opposite angles to
the normal.  Librations around an axis in this direction would greatly
affect overlap  of adjacent molecules and therefore the rate of polaron
motion. This appears to be the situation in at least one of the
structures attributed to $\alpha$-6T.\cite{hor3}
In such a case the libration
would have by far the strongest effect on the hopping and the effect
of the other phonons could be considered to be included in the value of
$g_i$, which is determined by comparison with experiment in any case.

As has been shown by many authors, in the approximation
that only one phonon branch interacts with the carriers the thermally
averaged bandwidth for the $i$th direction is\cite{comm2}
\begin{equation}
W_T=4|V_i|\exp[-g_i^2\coth(\beta\hbar\omega_0/2)]=4|V_i|\exp[-g_i^2(m+
1/2)]\;,
\end{equation}
where $m=[\exp(\beta\hbar\omega_0)-1]^{-1}$, the number of
phonons in the mode with frequency $\omega_0$. It is apparent that
$W_T$ decreases monotonically as the temperature increases from zero,
rapidly when $k_BT/\hbar\omega_0  > 1/2$.  Thus the mobility,
which is proportional to $W_T^2$, decreases with temperature due
to this factor, rapidly at high temperatures.  A second factor depending
on $T$ is $\mbox{csch}(\beta\hbar\omega_0/2)$ in the integrand.  This
factor is proportional to $m$, the phonon abundance.  In the low
temperature limit this factor makes the argument of $I_0$ small and
therefore the integral small.  As $T$ increases this factor increases
and the integral increases, quite rapidly for $k_BT>\hbar\omega_0$.
Crudely it can be thought that the integral represents the hopping
probability.  Thus as $T$ increases from zero the librons play the
dual role of decreasing the coherent or bandlike motion through
the decrease in bandwidth and of increasing the hopping rate.  How
strong these effects are depends on $g_i$.

Before comparing these predictions with the data of Torsi {\em et al}.
shown in Fig.\ 1, we note that their mobility values do not show the
effect of trapping; their measurements have  been made for gate voltages
past the threshold value required to obtain a carrier concentration
large enough to fill the traps. Also they have shown that grain
boundaries do not dominate the transport, at least above the minimum
in $\mu $ at $\sim 45$\ K. Increase in grain size by an order of magnitude
did not improve the room temperature mobility.\cite{to} From these facts
Torsi {\em et al}. conclude that their FET mobility data represent
intrinsic behavior of $\alpha$-6T.  Comparing the data of Fig.\ 1 with
the theory discussed above, we see above the temperature of
the minimum, $T_L$,  a region in which
$\mu$ increases rapidly with increasing $T$, consistent with hopping.
Below $T_L$ there is a region in which $\mu$ decreases rapidly
with increasing $T$, as expected for the coherent motion.  The fact that
$\mu$ does not increase without limit with decreasing $T$, as predicted
by Eq.\ (\ref{mu}), must be  due to another process, perhaps
boundary  scattering, determining $\mu$ in the low temperature limit.

To fit the data of Fig.\ 1 with Eq.\ (\ref{mu}) we took $a_i=0.38$\ nm,
the closest distance between adjacent molecules along the transport
direction.\cite{to} The widths of the phonon band, not known for
$\alpha$-6T, were taken from a libration  of similar frequency
which determines the polaron transport in naphthalene,\cite{ken}
specifically $\hbar\Delta_1=0.70$\ meV, $\hbar\Delta_2=1.40$\ meV and
$\hbar\Delta_3=0.24$\ meV. In any case $\mu_{ii}$ is not sensitive to the
$\Delta_i$ values. The quantities $|V_i|$, $g_i$, $\hbar\omega_0$
and $\alpha$ were treated as parameters.  The fit shown in Fig.\ 1 was
obtained for $|V_i|=14$\ meV, $g_i^2=11.5$, $\hbar\omega_0=11.2$\ meV
and $\alpha=2\times 10^{-5}$\ ps$^{-1}$. The first three parameters are
quite close to those obtained by Torsi {\em et al}. using Holstein'
s results.

It is of interest to compare these results with those obtained
by KADD in fitting Eq.\ (\ref{mu}) to the data for naphthalene.
From 30\ K, the lowest temperature for which there are data, to
$\sim 100$\ K $\mu$ for naphthalene decreases fairly rapidly
with increasing $T$, although not as rapidly as for $\alpha$-6T.
Above 100\ K $\mu$ continues to decrease with increasing $T$ but
the increase is much less rapid in two of the three directions shown.
These results are consistent with Eq.\ (\ref{mu}) with a somewhat larger
$\hbar\omega_0$ but more importantly a smaller value of $g_i$ as
compared with $\alpha$-6T.
Decreasing the electron-phonon coupling decreases both the rate
of bandwidth decrease and the rate at which hopping grows with
increasing $T$.  The value of $g_i^2$ for naphthalene, varying from
2.6 to 3.5 over the three directions, is small enough that
the increase in the integral is not large enough to overcome the
decrease in bandwidth and create a positive slope for mobility vs.
temperature.  Another contrasting feature of the naphthalene data is
that the changes in slope of $\mu$ vs. $T$ are much more
gradual.  This is apparently a result of $\alpha$, the disorder
parameter, being larger in naphthalene, 0.259 vs
$2\times 10^{-5}$\ ps$^{-1}$. In fact, the later value for $\alpha$
is surprisingly small. We speculate that this may be a
result of the holes  being confined to within 1 or 2 molecular layers
of the interface\cite{do} so that only a very small part of the
$\alpha$-6T layer is sampled, that  part being more ordered
perhaps because of the proximity of the interface.  Of course, it
is not possible to be certain of the parameters because there may
still be some differences between $\mu_{FET}$ and the drift
mobility, particularly below 45\ K.

If higher mobility is the requirement for improved organic FET
performance, it is clear from Eq. (4) that for molecular crystals
desirable properties are lower electron-phonon coupling and lower
frequency of the interacting phonons.  There is a limit to how far
one can go in this direction, however, because polaron theory is no
longer valid if $g^2\hbar\omega_0/W<1$. From what was said
earlier, naphthalene is superior to $\alpha$-6T, its room temperature
mobility ranging  from 0.3 to 0.8\ cm$^2$/Vs in the different
crystallographic directions.\cite{ken}

We acknowledge the support of the National Science Foundation
under Science and Technology Center grant CHE912001.

\begin{figure}
\caption{$\mu_{FET}$ vs. $T$. Squares give experimental data
of Torsi {\em et al.},$^{7}$ solid curve is the theoretical
fit described in the text.}
\end{figure}

\end{document}